\documentstyle[prl,tighten,aps,epsf,amstex,multicol,palatino,bbm]{revtex}

\begin{document}

\newcommand{\proofend}{\hfill\rule{2mm}{2mm} \\\bigskip }

\centerline{\Large\bfseries
On the quantification of entanglement}
\smallskip

\centerline{\Large\bfseries
in infinite-dimensional quantum systems}

\bigskip\bigskip

\centerline{Jens Eisert$^{1}$, Christoph
Simon$^{2}$, and Martin B. Plenio$^{1}$}

\medskip

\centerline{\small 1
QOLS, Imperial College of Science, Technology and Medicine,
London, SW7 2BW, UK}

\centerline{\small 2
Clarendon Laboratory, University of
Oxford, Oxford OX1 3PU, UK}

\bigskip

\begin{center}
\small
\begin{minipage}{1.0\textwidth}
We investigate entanglement measures in the infinite-dimensional
regime. First, we discuss the peculiarities that may occur if the
Hilbert space of a bi-partite system is infinite-dimensional, most
notably the fact that the set of states with infinite entropy of
entanglement is trace-norm dense in state space, implying that in
any neighbourhood of every product state lies an arbitrarily
strongly entangled state. The starting point for a clarification
of this counterintuitive property is the observation that if one
imposes the natural and physically reasonable constraint that the
mean energy is bounded from above, then the entropy of
entanglement becomes a trace-norm continuous functional. The
considerations will then be extended to the asymptotic limit, and
we will prove some asymptotic continuity properties. We proceed by
investigating the entanglement of formation and the relative
entropy of entanglement in the infinite-dimensional setting.
Finally, we show that the set of entangled states is still
trace-norm dense in state space, even under the constraint of a
finite mean energy.\\

\noindent
PACS-numbers: 03.67.-a, 03.65.Bz
\end{minipage}
\end{center}

\bigskip

\hrule\hrule
\bigskip

\section{Introduction}

When A.\ Einstein, B.\ Podolski, and N.\ Rosen published their
seminal paper on the question whether the quantum mechanical
description of physical reality could be considered complete, they
formulated their gedanken experiment in terms of two quantum
systems that are entangled through their position and momentum. In
the well-known reformulation by D.\ Bohm the entangled quantum
systems are replaced by two spin-1/2-particles, each of which is
supplemented with a two-dimensional Hilbert space \cite{Bohm}. In
this formulation the physical situation can be discussed without
the need to take the technicalities of infinite-dimensional spaces
into account. Most of the theory of quantum information has in
fact been developed for finite-dimensional settings, the qubit --
the two-level system -- being the prototypical elementary quantum
system in this context \cite{Intro1}.
Only quite recently, several contributions in the field addressed
issues of quantum information in infinite-dimen\-sio\-nal settings
\cite{Clifton,Majewski,Simon,OurGaussian,Lloyd}.
Experiments have been performed with quantum optical systems where
not the polarisation degrees of freedom have been employed, but
the ones corresponding to the canonical coordinates of the field
modes. Influenced by such experiments the theory of entanglement
has been extended to systems with infinite-dimensional Hilbert
spaces, most notably for so-called Gaussian quantum states,  which
form an important class of states from a practical point of view
\cite{Simon,OurGaussian}. 
In particular, the question of separability and distillability of
Gaussian states of multi-mode systems is essentially solved.
Concerning the theory of entanglement of non-Gaussian states,
however, not so much is known to date.

It is the purpose of this paper to comment on some
peculiarities that may occur in the infinite-dimensional
setting when quantifying the degree of entanglement,
and on ways to retain meaningful measures
of entanglement. The starting point is the following
observation: on a bi-partite infinite-dimensional Hilbert space
there always exist pure states which are arbitrarily close
to a pure product state - as quantified by the trace-norm
difference --
but they exhibit an entropy of
entanglement which is infinite. In this sense the assignment
of the value zero of the entanglement of the pure
product state is not entirely unambigous. A similar
situation can occur in the mixed-state domain.
The key insight leading to a possible
resolution of this problem
is derived from an investigation of the {\it mean energy}
of the involved quantum states:
If one imposes the natural
requirement that the mean energy
of the involved states is bounded from above,
then one retains meaningful measures of entanglement.
In other words, although there are indeed sequences of
pure states that converge
to a certain pure state in trace-norm while the
entropy of entanglement is divergent,
their energy must necessarily diverge.

The scope of this paper is quite modest:
We will neither characterize entanglement measures
through certain natural axioms, nor try to
equip the known entanglement measures with
a clearcut operational
interpretation in the infinite-dimensional
setting. Instead, this paper presents
a collection of clarifying statements concerning the
continuity of entanglement measures. We will
discuss the trace-norm continuity of the entropy
of entanglement and an asymptotic continuity property.
In addition, we will study the relative entropy of
entanglement and the entanglement of formation.
In the last section we will consider the question of
the existence of a separable neighbourhood of some mixed state
under a constraint on the mean energy.

\section{Notation}

We start by clarifying the notation that will be used in
this paper. We will consider
the situation where the composite system consists of
two parts $A$ and $B$ (each of which
having finitely many degrees of freedom)
such that the Hilbert space of
the joint system can be written as
${\cal H}={\cal H}_A\otimes {\cal H}_B$.
Throughout the paper
we will assume that
$\text{dim}[{\cal H}_A]= \infty$
and
$\text{dim}[{\cal H}_B]= \infty$.
The notation
${\cal S}({\cal H})$
will be used for the
set of density operators on ${\cal H}$, that is, the set of
positive trace-class
operators with trace one. In the same way
the set ${\cal S} ({\cal H}_A)$ is
defined as the state space on ${\cal H}_A$.
The symbol
$\|.\|$ will be reserved for
the trace-norm, which is defined as
$\|A\|=\text{tr}[| A|]=
\text{tr}[(A^\dagger A)^{1/2}]$
for operators $A$, while the standard operator
norm will be written as $|||.|||$.
%
%
%
For our purposes it will also be necessary to
impose a certain requirement on the Hamiltonian
$H=H_{A}\otimes 1 + 1\otimes H_{B}$ of the bi-partite
system. It will be demanded that $H$ has a discrete
spectrum, such that the spectral decomposition of
$H_A$ and $H_B$
reads as
\begin{equation}
H_A= \sum_{n=0}^{\infty} \epsilon_A^{(n)}
\pi_{A}^{(n)}\,\,\text{ and }\,\,
H_B= \sum_{n=0}^{\infty} \epsilon_B^{(n)}
\pi_{B}^{(n)},
\end{equation}
respectively, where $\epsilon_{A}^{(i)}\leq
\epsilon_{A}^{(j)}$ and
$\epsilon_{B}^{(i)}\leq
\epsilon_{B}^{(j)}$ for $i <j$.
$\pi_{A}^{(n)}$ are the
projectors on the
one-dimensional spaces spanned by the mutually orthogonal
$\phi_A^{(n)}\in{\cal H}_A$, and analogously
for $\pi_{B}^{(n)}$.
We will always
assume
that the Hamiltonian $H$
of the system has the property
that
\begin{equation}
    \text{tr}[e^{-\beta H}]<\infty
\end{equation}
for all $\beta>0$. This is a natural
assumption: it is simply required that the
Gibbs state exists for all inverse temperatures of
the system. In particular, it means that there can be no
limit points in the spectrum of $H$, which in turn implies
that no eigenvalue can be infinitely degenerate.

\section{Entropy of Entanglement}

Entanglement measures give an account of the degree
of entanglement of a given quantum state. They are
functionals mapping state space on the set of non-negative
real numbers; the larger the number is, the more entangled
is the quantum state.
For pure states
we will investigate
the
entropy of entanglement $E:{\cal S}({\cal H})\longrightarrow{\mathbbm{R}}^+$
(or reduced entropy), defined as
\begin{equation}
    E(\rho):=(S\circ \Phi_B)(\rho).
\end{equation}
Here, $S:{\cal S}({\cal H}_A)\longrightarrow {\mathbbm{R}}^+$
denotes the von-Neumann entropy, which is
defined as $S(\rho)=-\text{tr}[\rho\log_2(\rho)]$,
and $\Phi_B:{\cal S}({\cal H})\longrightarrow
{\cal S}({\cal H}_A)$ is the partial trace with respect to system $B$.
The entropy of entanglement quantifies to what extent
the state departs from a product state.
As has already been pointed out in the introduction,
the entropy of entanglement may be very different from zero
for states which are very `close' to pure product states:\\


\noindent
{\bf Example 1. }
Let
    $\sigma_0= |\psi_0\rangle\langle\psi_0|$,
$\psi_0= \phi_{A}^{0}\otimes \phi_{B}^{0}$,
be the ground state of the
bi-partite system,
and let
$\{\sigma_k\}_{k=1}^{\infty}$
be a sequence of pure states
$\sigma_k=|\psi_k\rangle\langle\psi_k|$,
    where
\begin{eqnarray}
    \psi_k : =
    (1-\delta_k)^{1/2}
    \psi_0+(\delta_k/k)^{1/2}
    \sum_{n=1}^k \phi_{A}^{(n)}\otimes \phi_{B}^{(n)}.
\end{eqnarray}
and
$\delta_k := 1/
\log (k)$. In fact, the series $\{\sigma_k\}_{k=1}^{\infty}$
converges to $\sigma_0$ in trace-norm, i.e., $\lim_{k\rightarrow \infty}
\| \sigma_k - \sigma_0\|=0$. However,
\begin{equation}
    \lim_{k\rightarrow \infty}E(\sigma_k)=1,
\end{equation}
whereas $E(\sigma_0)=0$, since $\sigma_0$
it is a product state. \\

That is to say, $E=S\circ \Phi_B$ is clearly
not continuous in $\sigma_0$:
There are states
which are arbitrarily close
to the product ground state with respect to the trace-norm,
but they have
an entropy of entanglement that
approaches a different value.
As the trace-norm difference is an upper bound for
differences in expectation values of all contractions,
it is clear that the smaller the trace-norm difference
of a pair of states is, the more difficult is it to
distinguish the states by operational means.
The situation is actually even more surprising,
as the set of states which have infinite von-Neumann
entropy is trace-norm dense in state space \cite{Wehrl}.
As can be easily verified, the set of pure states
with infinite entropy of entanglement is also trace-norm
dense in the set of all pure states:\\

\noindent
{\bf Proposition 2. } {\it
For all $\psi\in{\cal H}$ and all
$\varepsilon>0$ there exist a
vector $\phi\in{\cal H}$ such that $\|\, |\psi\rangle\langle\psi| -
|\phi\rangle\langle\phi|\,  \|<\varepsilon$
and $(S\circ \Phi_B)(|\phi\rangle\langle\phi|)=\infty$.}\\
%

{\it Proof.}
According to the Schmidt decomposition
(which can be applied in this infinite-dimensional setting)
there exists an orthonormal basis
$\{\psi_{A}^{(n)} \}_{n=1}^{\infty}$
of ${\cal H}_{A}$ and a basis
$\{\psi_{B}^{(n)} \}_{n=1}^{\infty}$
of ${\cal H}_{B}$ such that
$\psi\in{\cal H}$ can be written in the
form
\begin{equation}
    \psi=\sum_{n=1}^{\infty}
(p^{(n)})^{1/2}\psi_{A}^{(n)} \otimes \psi_{B}^{(n)} ,
\end{equation}
where $\{p^{(n)} \}_{n=0}^{\infty}$
forms a probability distribution.
For each $n\in{\mathbbm{N}}$, define
a sequence $\{q_{k}^{(n)}\}_{k=1}^{\infty}$
through $q_{k}^{(1)}= p_{k}$, and
\begin{equation}
q_{k}^{(n)}:= \frac{p^{(n)} + 1/(k n \log_2(n)^{2})}{\delta_{k}},\,\,\,\,\,
n=2,3,\ldots,
\end{equation}
where $\delta_{k}>0$ is chosen in such
a way that $\{q_{k}^{(n)}\}_{n=1}^{\infty}$
is also a probability distribution. This
gives rise to a sequence of state vectors $\psi_{k}\in{\cal H}$
defined as
\begin{equation}
    \psi_{k}:= \sum_{n=1}^{\infty}
(q_{k}^{(n)})^{1/2}\psi_{A}^{(n)} \otimes \psi_{B}^{(n)}.
\end{equation}
The sequence
$\{p^{(n)} \}_{n=0}^{\infty}$ is convergent;
on using the fact that
$f:[0,1]\longrightarrow {\mathbbm{R}}^{+}$
defined as $f(x)=-x \log_2(x)$ is
monotone increasing in $[0,\varepsilon]$ for
some $\varepsilon>0$,
one can show that
$(S\circ \Phi_{B})(|\psi_{k}\rangle\langle\psi_{k}|)=\infty$
for all $k\in{\mathbbm{N}}$. However,
\begin{equation}
    \lim_{k\rightarrow\infty}
    \|\, |\psi\rangle\langle\psi| -
    |\psi_{k}\rangle\langle\psi_{k} |\,  \|=
    \lim_{k\rightarrow\infty}
    \sum_{n=1}^{\infty}
    | p^{(n)}- q_{k}^{(n)}|=0.
\end{equation}
\proofend

The key observation in a possible clarification of this issue
concerns the  mean energy of the states. In the above example,
the mean energy $\text{tr}[\sigma_k H]$
grows beyond all bounds.
After all, the energy that can possibly
be invested in the preparation of a quantum
state is in all instances limited. This is no accident,
and we will see that in sequence of states $\{\sigma_k\}_{k=1}^\infty$
that converges in trace-norm to some state $\sigma$
such that
the sequence of the entropies of entanglement
of $\sigma_k$
does not converge to the entropy of entanglement of
$\sigma$, the
mean energy necessarily diverges.

Let ${\cal S}_M({\cal H})\subset{\cal S}({\cal H}) $
for a given number $M>0$ and for a given Hamiltonian $H$
be the set of states with a mean energy of at most $M$,
\begin{equation}
    {\cal S}_M({\cal H}):= \{ \rho\in {\cal S}({\cal H})|
    \text{tr}[\rho H] < M\}.
\end{equation}
Although this set is nowhere dense in state space  -- as $H$ is
an unbounded operator according to the previous assumptions --
it is a reasonable subset of the state space: it simply reflects
the natural requirement
that the mean energy is bounded from above.
On this set, it turns out, several
entanglement measures do
not exhibit the above pathologies.
We start with the trace-norm continuity of the
entropy of entanglement, which is a corollary of a theorem
in Ref.\ \cite{Wehrl} due to A.\ Wehrl. \\

\noindent
{\bf Proposition 3. } {\it Let $M>0$,
let $\sigma\in {\cal S}_M({\cal H})$,
and let $\{\sigma_k\}_{k=1}^{\infty}$
be a sequence of states $\sigma_k\in {\cal S}_{M}({\cal
H})$
satisfying $\sigma_k\longrightarrow \sigma$ in
trace-norm. Then
\begin{equation}
    \lim_{k\rightarrow\infty} |E(\sigma)- E(\sigma_k)|
    =0,
\end{equation}
that is, the entropy of entanglement
$E:{\cal S}_M({\cal H})
\longrightarrow {\mathbbm{R}}^+$
is a trace-norm continuous functional.} \\

{\it Proof. }\/
This statement is a
consequence of a statement concerning
the continuity of the von-Neumann entropy under
an appropriate constraint of the energy \cite{Wehrl,Wehrl2}:
If $\{\omega_k\}_{k=1}^\infty$ is a sequence
of states taken from ${\cal S}_M({\cal K})$
satisfying $\omega_k\rightarrow\omega$ in trace-norm
for some state $\omega\in {\cal S}_M({\cal K})$, together
with the above
assumptions concerning the spectrum of $H$,
then
\begin{equation}
\lim_{k\rightarrow \infty} S(\omega_k)=S(\omega).
\end{equation}
The states $\omega_k:=
\Phi_B(\sigma_k)$ with ${\cal K}:={\cal H}_A$
form such a sequence, since
\begin{equation}
    \|\Phi_B(\sigma_k)-\Phi_B(\sigma)\|\leq \| \sigma_k-\sigma\|
\end{equation}
for all $k\in {\mathbbm{N}}$ by the contraction property of the trace-norm
under completely positive maps, and since
\begin{equation}
\text{tr}[H_{A} \Phi_B(\sigma_k)] \leq
\text{tr}[H \sigma_k] < M.
\end{equation}
Hence,
$\lim_{k\rightarrow \infty}
|(S\circ \Phi_B)(\sigma_k)- (S\circ \Phi_B)(\sigma)|= 0$
if $\lim_{k\rightarrow \infty}
    \|\sigma_k-\sigma\|=0$.
\proofend

Therefore, if one introduces the natural and physically reasonable
restriction that the mean energy must be bounded from above, 
then
the
entropy of entanglement becomes continuous, and a situation as in
Example 1 or in Proposition 2 is excluded through this restriction
\cite{Zyk} . Also, the entropy of entanglement is in all instances
finite for pure states taken from ${\cal S}_{M}({\cal H})$ for
$M>0$.
The statement of Proposition 3
alone does not, however, guarantee that the entropy of
entanglement has appropriate
continuity properties to describe entanglement manipulation
in the asymptotic limit
\cite{Bennett1,Hayden,DonaldCont,Asympt,Unique},
in the sense of
the limit
of infinitely many
identically prepared
quantum systems.
In particular, in order to give the entropy of entanglement
an interpretation in terms of optimal achievable
conversion rates in distillation-type protocols
\cite{Bennett1}, one would aim at finding
the following continuity property:
Take a large number $m$
of identically prepared quantum systems in a state $\rho$.
In a distillation procedure one would like
to convert the state $\rho^{\otimes m}$
of these systems
into a product state $\sigma^{\otimes n}$ consisting
of a number of highly entangled states $\sigma$
of finite-dimensional quantum systems.
For any finite number of  identically prepared
systems
this procedure may not be possible in an optimal manner.
One would then obtain a state
$\sigma_n$, which is similar to yet not identical with
$\sigma^{\otimes n}$. One may
nevertheless require that
the transformation is optimal in the asymptotic
limit $n\longrightarrow\infty$. Roughly speaking,
the measure of entanglement should then assign
a value to $\sigma_n$ which is very similar
to the value for $\sigma^{\otimes n}$, such that the
difference per copy becomes negligible
in the asymptotic limit.
Note that
none of the mixed states $\sigma_{n}$
are required to have a finite-dimensional carrier.
It is the purpose of the subsequent proposition to show
that under
an appropriate restriction on the mean energies, the
entropy of entanglement exhibits
an asymptotic continuity property \cite{asy}.\\

%

\noindent
{\bf Proposition 4. } {\it
Let $\sigma\in {\cal S}_M({\cal H})$, $M>0$,
be a pure state that is supported
on a finite-dimensional
subspace of ${\cal S}({\cal H})$,
and let $\{\sigma_n\}_{n=1}^{\infty}$, $\sigma_n\in {\cal S}_{n M}({\cal
H}^{\otimes n})$,
be a sequence of
states satisfying
\begin{equation}
    \lim_{n\rightarrow\infty}
    \|\sigma_n -\sigma^{\otimes n}\|=0.
\end{equation}
Then
\begin{equation}
    \lim_{n\rightarrow\infty}
    \frac{|E(\sigma^{\otimes n})
    - E(\sigma_n)|}{n}
    =0.\label{Asy}
\end{equation}
}

The subsequent lemma will prepare the proof.
The
property of the
von-Neumann entropy
that will be investigated
reminds of the lower
semi-continuity property \cite{Ohya}.
In our case, however, we consider a
sequence of states
which are
defined on a sequence of Hilbert spaces
of increasing dimension.\\

\noindent
{\bf Lemma 5. } {\it Let $\omega$ be a state
that is supported on a finite-dimensional subspace of
${\cal S}({\cal H}_A)$,
and let $\{\omega_n\}_{n=1}^{\infty}$,
$\omega_n\in {\cal S}({\cal H}_A^{\otimes n})$,
be a sequence of states satisfying
\begin{equation}\nonumber
    \lim_{n\rightarrow \infty}
\|\omega_n -\omega^{\otimes n}\|=0.
\end{equation}
Then
\begin{equation}\label{StartingPoint}
    S(\omega )
    \leq
    \liminf_{n\rightarrow\infty}
     \frac{1}{n}
    S(\omega_n).
\end{equation}
}

{\it Proof. }
Let $\pi$ be the projection operator on the support of $\omega$,
and let
\begin{equation}
    \eta_n:=
\pi^{\otimes n} \omega_n \pi^{\otimes n},\,\,
\lambda_n:=  \| \eta_n\|.
\end{equation}
The trace-norm distance is non-increasing under
trace-preserving completety positive maps, and
in particular, under pinchings \cite{Bhatia}.
Therefore,
\begin{eqnarray}
    \|\omega^{\otimes n} - \eta_n \| + (1-\lambda_n)
    &\leq & \|  \omega^{\otimes n} -\omega_n\|.
\end{eqnarray}
Hence, if
$\lim_{n\rightarrow \infty}
    \|
    \omega^{\otimes n} -\omega_n\|=0$ holds,
then also $\lim_{n\rightarrow \infty}
    \|
\omega^{\otimes n} -\eta_n\|=0$,
and $\lim_{n\rightarrow \infty } \lambda_n=1$.
In turn, if $\lim_{n\rightarrow \infty}
    \| \omega^{\otimes n}-\eta_n\| = 0$, then
\begin{equation}
\lim_{n\rightarrow \infty}
    \| \lambda_n\omega^{\otimes n}-\eta_n\| = 0,
\end{equation}
as
\begin{equation}\label{AlsoNeeded}
\|\lambda_n\omega^{\otimes n}-\eta_n\|
\leq | \lambda_n -1| + \| \omega^{\otimes n}-\eta_n\|.
\end{equation}
The triangle inequality yields
\begin{eqnarray}
 |S(\omega^{\otimes n})-S(\eta_n )|/n
&\leq& |S(\omega^{\otimes n}) - S( \eta_n/\lambda_n)|/n\nonumber\\
    &+&
     | 1- 1/\lambda_n| S(\eta_n)/n -
    \log_2(\lambda_n)/(n\lambda_n).\label{Tri}
\end{eqnarray}
The second term on the right hand side in
Eq.\ (\ref{Tri}) certainly vanishes in the limit
$n\rightarrow \infty$, as $ S(\eta_n)/n\leq C$
for all $n\in{\mathbbm{N}}$ for some appropriately
chosen $C>0$.
By applying Fannes' inequality \cite{FannesIn}
on the first term and
by making use of Eq.\ (\ref{AlsoNeeded})
one can conclude that
\begin{equation}\label{Ergebnis1}
    \lim_{n\rightarrow \infty}\frac{|S(\omega^{\otimes n})-S(
\pi^{\otimes n} \omega_n \pi^{\otimes n} )|}{n} = 0
\end{equation}
if $\lim_{n\rightarrow \infty}\| \omega^{\otimes n} - \omega_n\|=
0$. Using the function $f$ defined as $f(x)=-x \log_2(x)$
one finds that
\begin{eqnarray}
    \liminf_{n\rightarrow \infty}
    \frac{1}{n} S(\omega_n)&\geq&
    \liminf_{n\rightarrow \infty}
    \frac{1}{n} \text{tr}[\pi^{\otimes n}
    f(\omega_n)]\nonumber
    =S(\omega),
\end{eqnarray}
which is the statement of the lemma. \proofend

{\it Proof of Proposition 4:} The first step
of the proof can be performed
just as in Lemma 5.
Let $\xi\in {\cal S}_M({\cal H})$,
let $\omega$ be a state
that is supported on a finite-dimensional
subspace of ${\cal S}_M({\cal H})$,
and let $\{\omega_n\}_{n=1}^{\infty}$,
$\omega_n\in {\cal S}_{n M}({\cal H}_A^{\otimes n})$,
be a sequence of states satisfying $\lim_{n\rightarrow \infty}
\|\omega_n -\omega^{\otimes n}\|=0$.
Then
\begin{equation}\label{StartingPoint}
    S(\xi|| \omega )
    \leq
    \liminf_{n\rightarrow\infty} \frac{1}{n}
    S(\xi^{\otimes n} ||
    \omega_n)
\end{equation}
holds.
The validity of Eq.\ (\ref{StartingPoint})
can be seen as follows:
the relative entropy
of $\xi^{\otimes n}$ with respect to $\omega^{\otimes n}$
can be written as \cite{Ohya,Wehrl}
\begin{eqnarray}
 S(\xi^{\otimes n}|| \omega^{\otimes n})&=&
\sup_{\mu\in [0,1]}\sup_{\pi_n}
\text{tr}[\pi_n(
    f( \mu \xi^{\otimes n}+ (1-\mu) \omega^{\otimes n})
    \nonumber\\
    &-& \mu f(\xi^{\otimes n}) -
    (1-\mu) f(\omega^{\otimes n}))
    ],
\end{eqnarray}
where the supremum is taken over all
finite projection operators $\pi_n$
on ${\cal H}_A^{\otimes n}$.
So just as in Lemma 5,
applying the triangle inequality and Fannes'
inequality several times yields Eq.\ (\ref{StartingPoint}).
In particular, one has to make use of
\begin{eqnarray}
    &&\frac{|S(\mu \xi^{\otimes n} + (1-\mu) \omega^{\otimes n})
    -S(\mu \xi^{\otimes n} + (1-\mu) \eta_n)|}{n}\nonumber\\
    &\leq &
    \frac{|S(\mu \xi^{\otimes n} + (1-\mu) \omega^{\otimes n})
    -S(\mu \xi^{\otimes n} + (1-\mu) \eta_n/\lambda_n)|}{n}\nonumber\\
    &+&
    \frac{|S(\mu \xi^{\otimes n} + (1-\mu) \eta_n/\lambda_n)-
    S(\mu \xi^{\otimes n} + (1-\mu) \eta_n)|
    }{n}.
\end{eqnarray}
Eq.\ (\ref{StartingPoint})
is now the starting point of an argument
along the line of the argument of
Ref.\ \cite{Wehrl}:
Since the free energy can be expressed in terms
of the relative entropy according to
\begin{eqnarray}\nonumber
    \frac{1}{n}F(\omega_n, \beta, H^{\otimes n})
    &=&(
    S(\sigma_\beta^{\otimes n}|| \omega_n)/n -
    \log_2(\text{tr}[
    e^{-\beta H}
    ])/\beta,
\end{eqnarray}
it has the property
\begin{equation}
    F(\omega, \beta, H)
    \leq \liminf_{n\rightarrow \infty}\frac{1}{n}
    F(\omega_n, \beta, H^{\otimes n}),
\end{equation}
implying that
\begin{equation}
    \beta
    \text{tr}[\omega H]-
    S(\omega) \leq
    \liminf_{n\rightarrow \infty}
    \frac{1}{n}
    \bigl(
    \beta
    \text{tr}[
    \omega_n H^{\otimes n}
    ]- S(\omega_n)
    \bigr),
\end{equation}
and, therefore,
\begin{eqnarray}\nonumber
    -S(\omega)
    &\leq& \liminf_{n\rightarrow \infty}
    (-S(\omega_n)/n ) \nonumber\\
    &+&
    \beta
    \limsup_{n\rightarrow
    \infty}
    |
    \text{tr}[
    \omega_n H^{\otimes n}
    ]
    |/n+
    \beta
    \text{tr}
    [\omega H]\label{WehAs}
\end{eqnarray}
for all $\beta>0$. Again,
the sum of the last two terms
of Eq.\ (\ref{WehAs})
is bounded from above by $2\beta M$,
providing the inequality
\begin{equation}
    -S(\omega)\leq
    \liminf_{n\rightarrow \infty}
    (-S(\omega_n)/n).\label{Almost}
\end{equation}
Eq.\ (\ref{Almost}) and
the statement of Lemma 4
then imply that
$S(\omega)=\lim_{n\rightarrow \infty}
    S(\omega_n)$.
Finally, one can argue as in Proposition 3, by taking
$\omega:= \phi_B(\sigma)$
and $\omega_n:=  \Phi_B(\sigma_{n})$.
This gives rise to the asymptotic continuity property
stated in Eq.\ (16).
\proofend

This asymptotic continuity property
concerned the entropy of entanglement.
A similar property would be desirable
for sequences of states that are possibly
mixed states. In this context it is
of interest
to see whether typical measures for mixed states
give, loosely speaking,
the appropriate value when they are evaluated
for pure states. More precisely, one may ask whether
entanglement measures are trace-norm continuous in
pure states, and whether they exhibit an asymptotic
continuity property.
We will discuss these questions for
the relative entropy of entanglement \cite{Quant1}
and the entanglement of formation \cite{Bennett1},
starting with the latter one:

For a given mixed
state $\sigma\in {\cal S}_M({\cal H})$, $M>0$,
there exist
(uncountably many) sequences $\{p^{(i)}\}_{i=1}^\infty$
of positive numbers forming a probability distribution, $p^{(1)}\geq p^{(2)}\geq...$,
and sequences $\{\psi^{(i)}\}_{i=1}^\infty$ of
state vectors $\psi^{(i)} \in {\cal H}$
such that
\begin{equation}
    \sigma= \sum_i p_i |\psi^{(i)}\rangle\langle\psi^{(i)}|.
\end{equation}
The pair $(\{p^{(i)}\}_{i=1}^\infty, \{\psi^{(i)}\}_{i=1}^\infty)$
will be called decomposition of $\sigma$. As for a
finite-dimen\-sio\-nal
Hilbert space one may define the entanglement of formation as
\begin{equation}
    E_F(\sigma)= \inf \sum_i p_i (S\circ \Phi_B)
    (|\psi^{(i)}\rangle\langle\psi^{(i)}|),
\end{equation}
where the infimum is understood with respect to all decompositions
of $\sigma$. In the case of a finite-dimensional Hilbert space the
infimum is always attained, and by virtue of Caratheodory's theorem
one can find an upper bound for the required number of terms in a a
decomposition of the state.
It is worth noting that, using
a different language, a decomposition of a state $\sigma$
can also be represented by probability measures $\mu_\sigma$
on state space with the
barycenter $b(\mu_\sigma)=\sigma$. The entanglement of formation can
then be defined via a minimization of the mean
local von-Neumann entropy
with respect to all
probability measures with the same barycenter.
This approach has been pursued in
Ref.\ \cite{Majewski}, where the entanglement of
formation has been studied in the case of a bi-partite
system, with of the systems being finite-dimensional.
The main difference of the situation in Ref.\
\cite{Majewski} and in the present paper is that in
the case of one finite-dimensional subsystem
the trace-norm continuity of the von-Neumann entropy
is available.\\

\noindent {\bf Proposition 6.}
{\it Let $M>0$, let $\sigma\in{\cal S}_M({\cal H})$,
$\sigma=|\psi\rangle\langle \psi|$,
be a pure state, and let $\{\sigma_k\}_{k=1}^\infty$
a sequence of states $\sigma_k \in{\cal S}_M({\cal H})$
with $\sigma_k\longrightarrow \sigma$ in trace-norm.
Then
\begin{equation}
    \lim_{k\rightarrow\infty}
    |E_F(\sigma_k)- E_F(\sigma)|=0.
\end{equation}
}

{\it Proof.}
We start with proving the lower semi-continuity of
$E_F$ in $\sigma$.  Let  $\xi>0$ be a number satisfying
$E_F(\sigma_k)\leq \xi$ for all $k\in{\mathbbm{N}}$.
For all $\varepsilon>0$ there exists a decomposition
$(\{p_k^{(i)}\}_{i=1}^\infty, \{\psi_k^{(i)}\}_{i=1}^\infty)$
of each
$\sigma_k$
such that
\begin{equation}\label{lhs}
    \sum_i p_k^{(i)}
    (S\circ \Phi_B) (|\psi^{(i)}_k\rangle\langle\psi_k^{(i)}|)
    \leq \xi+\varepsilon.
\end{equation}
The fact that $\sigma_k\longrightarrow \sigma=|\psi\rangle\langle\psi|$
in trace-norm implies that there exists a sequence
of real numbers $\{f_k\}_{k=1}^\infty $
with $\lim_{k\rightarrow\infty} f_k=0$, and
such that
\begin{equation}
    \lim_{k\rightarrow\infty} \sum_{i} p_k^{(i)} \theta(
    f_k -
    \|\, |\psi^{(i)}_k\rangle\langle\psi_k^{(i)}| -
    \sigma\| )=1,
\end{equation}
where $\theta:{\mathbbm{R}}\longrightarrow \{0,1\}$
is the Heaviside function.
For each $k$ construct a sequence
 $\{q^{(i)}_k\}_{i=1}^\infty $ of real numbers
as
\begin{equation}
q^{(i)}_k=
\left\{
\begin{array}{ll}
p_k^{(i)}, &  \text{if
$f_k - \|\, |\psi^{(i)}_k\rangle\langle\psi_k^{(i)}| -
\sigma\|>0$, and}\\
q^{(i)}_k=0 & \text{otherwise.}
\end{array}
\right.
\end{equation}
Similarly, define a sequence
$\{\phi_k^{(i)}\}_{k=1}^\infty $
through $\phi_k=\psi_k$ if
$f_k - \|\, |\psi^{(i)}_k\rangle\langle\psi_k^{(i)}| -
\sigma\|>0$, and $\phi_k=\psi$ otherwise.
It follows that
\begin{equation}
\sum_i q_k^{(i)}
    (S\circ \Phi_B)
    (|\phi^{(i)}_k\rangle\langle\phi_k^{(i)}|)\leq \xi+\varepsilon,
\end{equation}
$\lim_{k\rightarrow\infty} \| \, |\phi^{(i)}_k\rangle\langle\phi_k^{(i)}|-\sigma\|=0$
for all $i$,
and $\lim_{k\rightarrow\infty} \sum_i q_k^{(i)}=1$.
Hence, due to the lower-semicontinuity of
the von-Neumann entropy we can conclude that also
$(S\circ \Phi_B)
(\sigma)\leq \xi+\varepsilon$. As $\varepsilon>0$ was
arbitrary, it follows that $E_F(\sigma)\leq \xi$, which means
that $E_F$ is lower semi-continuous in $\sigma$.
Note that the constraint on the mean energy was not needed in this step.

The second part of the proof will be concerned
with the upper semi-continuity in $\sigma$.
Let  $\xi\in{\mathbbm{R}}$ such that
$E_F(\sigma_k)\geq \xi$ for all $k\in{\mathbbm{N}}$.
Essentially, the proof goes again to a large extent
along the line of the argument
in Ref.\ \cite{Wehrl}, where additionally, we make use of the
convexity of the relative entropy functional.
Let $(\{p_k^{(i)}\}_{i=1}^\infty, \{\psi_k^{(i)}\}_{i=1}^\infty)$
be a decomposition of $\sigma_k$, and
let $\omega:= \Phi_B(\sigma)$,
$\omega_k:= \Phi_B(\sigma_k )$, and
$\omega_k^{(i)}:= \Phi_B( |\psi_k^{(i)}\rangle\langle \psi_k^{(i)}|)$.
On using both the lower semi-continuity and the convexity
of the relative entropy functional one obtains
\begin{equation}
    S(\omega||\sigma_\beta) \leq \liminf_{k\rightarrow\infty}
    S(\omega_k||\sigma_\beta) \leq
    \liminf_{k\rightarrow\infty}
    \sum_i p_k^{(i)} S( \omega_k^{(i)}
    ||\sigma_\beta).
\end{equation}
Therefore,
    $F(\omega,\beta,H_A) \leq
     \liminf_{k\rightarrow\infty}
    \sum_i p_k^{(i)} F(\omega_k^{(i)} ,\beta,H_A),$
which means that
\begin{equation}
\beta \text{tr}[\omega H_A]-S(\omega)
\leq
\liminf_{k\rightarrow\infty} \Bigl(\beta
\text{tr}[\omega_k H_A]-
\sum_i p^{(i)} S(\omega_k^{(i)} ) \Bigr)
\end{equation}
and
\begin{eqnarray}
    -S(\omega)
    \leq \liminf_{k\rightarrow \infty}
    \sum_i p^{(i)} (-S(\omega_k^{(i)}) )
    +
    \beta
    \limsup_{k\rightarrow
    \infty}
    |
    \text{tr}[
    \omega_k H_A
    ]
    |+
    \beta
    \text{tr}
    [\omega H_A].
\end{eqnarray}
Hence, we arrive at
\begin{equation}\label{Upside}
    -S(\omega)\leq
    \liminf_{k\rightarrow \infty}\sum_i p^{(i)}
    (-S(\omega_k^{(i)})).
\end{equation}
As $E_F(\sigma_k)\geq \xi$ for all $k\in{\mathbbm{N}}$,
and the above decomposition is not necesarily optimal
for $E_F(\sigma_k)$,
\begin{equation}
    \sum_i  p^{(i)}
(S\circ \Phi_B)(|\psi_k^{(i)}\rangle\langle \psi_k^{(i)}|)\geq \xi.
\end{equation}
The last step is to see that therefore,
$(S\circ \Phi_B)(\sigma)\geq \xi$, which means that
$E_F$ is also upper semi-continuous in $\sigma$.

\proofend

Therefore, we conclude, the entanglement of formation for
pure states can indeed by simply identified with the
entropy of entanglement on the set ${\cal S}_{M}({\cal H})$.
A similar argument applies again
on the asymptotic limit: if we have a series
of mixed states that converges in trace-norm to the
$n$-fold tensor product of a pure state with
a finite support, then, again, one can expect an
asymptotic continuity as in Proposition 4:\\

\noindent
{\bf Proposition 7.} {\it
Let $\sigma\in {\cal S}_M({\cal H})$, $M>0$,
be a pure state that is supported
on a finite-dimensional
subspace of ${\cal S}({\cal H})$,
and let $\{\sigma_n\}_{n=1}^{\infty}$,
$\sigma_n\in {\cal S}_{n M}({\cal
H}^{\otimes n})$, be a sequence of states satisfying
$\lim_{n\rightarrow\infty}
    \|\sigma_n -\sigma^{\otimes n}\|=0$.
Then
\begin{equation}
    \lim_{n\rightarrow\infty}
    \frac{|E_F(\sigma^{\otimes n})
    - E_F(\sigma_n)|}{n}
    =0.\label{Asy}
\end{equation}
}

{\it Proof.} One can proceed in the same way as
Proposition 6. Instead of the mere lower-semicontinuity
of the von-Neumann entropy one has to make use of
the statement of
Lemma 5.
\proofend

For general mixed states
the actual minimization that is necessary in order
to evaluate the entanglement of formation in
the infinite-dimensional setting
appears to be
quite intractable, and we do not consider
the entanglement of formation in this situation.

\section{Relative Entropy of Entanglement}

In the finite-dimensional setting
the relative entropy of entanglement
is defined as the minimal 'distance' of a given state
to an appropriately chosen set that is closed under
local quantum operations together
with classical communication (LOCC)
and which includes the identity \cite{Quant1}.
The distance is quantified
by means of the relative entropy functional.
Most typically, one chooses either the set of separable
states or the set of states with a positive partial transpose,
the set of PPT states. In case that
the Hilbert spaces are infinite-dimensional, one can
define the relative entropy of entanglement in the
same way -- with the exception that care is needed in
the definition of the notion of the set of separable states.
As in Refs.\ \cite{Werner89,Clifton}
we define the set of separable states
${\cal D}({\cal H})\subset {\cal S}({\cal H})$ as the set of states
for which there exists a sequence $\{\omega_k\}_{k=1}^\infty$,
$\omega_n\in  {\cal S}({\cal H})$,
such that $\omega_k\longrightarrow \omega$ in trace-norm, and
such that
each $\omega_k$ is of the form
\begin{equation}
    \omega_k= \sum_i p^{(n)}_i \eta_A^{(k,i)}\otimes \eta_B^{(k,i)}
\end{equation}
where $\eta_A^{(k,i)}\in {\cal S}({\cal H}_A)$,
$\eta_B^{(k,i)}\in {\cal S}({\cal H}_B)$ for all $i,k$,
and $\{p^{(k)}_i\}_{i=1}^\infty $
form probability distributions for all $k\in{\mathbbm{N}}$:
I.e., one requires
that the state $\omega$ can
be approximated in trace-norm
by convex combinations of products
\cite{Werner89,Clifton}, which in turn means that
${\cal D}({\cal H})$ is the
closed convex hull of the set of products (with respect
to the topology induced by the trace-norm).
Another reasonable set is the set
${\cal P}({\cal H})\subset {\cal S}({\cal H})$
of states that can be approximated in (trace-norm) by
PPT states,
that is, states $\omega\in {\cal S}({\cal H})$
for which their partial transpose $\omega^{T_A}$ is
again a state.
The relative entropy of entanglement is the
map $E_R: {\cal S}({\cal H})\longrightarrow {\mathbbm{R}}^+$
defined as
\begin{equation}
    E_R(\sigma)=\inf_{\rho\in {\cal D}({\cal H})}
    S(\sigma ||\rho),
\end{equation}
or, alternatively, with ${\cal D}({\cal H})$ being replaced
by ${\cal P}({\cal H})$.
Again, we are interested whether the relative entropy
of entanglement is trace-norm continuous in pure states.
Here, we will however show
the stronger property of
continuity on the whole state space:\\

\noindent
{\bf Proposition 8. } {\it The relative
entropy of entanglement $E_R:{\cal S}_M({\cal H})
\longrightarrow
{\mathbbm{R}}^+$ is trace-norm continuous.}\\

{\it Proof. }
Let $\{\sigma_k\}_{k=1}^{\infty}$ be a sequence of
states $\sigma_k\in {\cal S}_M({\cal H})$ for which
$\lim_{k\rightarrow\infty} \| \sigma_k - \sigma\|=0$.
Let
$\rho_k\in {\cal S}_M({\cal H})$ for each $\sigma_k$
be the state for which $E_R(\sigma_k)=S(\sigma_k\|\rho_k)$
(such a state exists, due to the compactness of the
set under consideration and the lower semi-continuity
of the relative entropy). Then
\begin{eqnarray}\label{eO}
    |E_R(\sigma)-E_R(\sigma_k)|&\leq&
    |S(\sigma)-S(\sigma_k)|\nonumber
    + |
    -\text{tr}[\sigma \log_2(\rho)]+
    \text{tr}[\sigma_k \log_2(\rho_k)]|.
\end{eqnarray}
As $\sigma,\sigma_k\in {\cal S}_M({\cal H})$,
$\lim_{n\rightarrow \infty}
|S(\sigma)-S(\sigma_k)|=0$.
%
The first term on the right hand side of Eq.\ (\ref{e0})
is bounded from above by
\begin{eqnarray}\label{all}
|-\text{tr}[\sigma \log_2(\rho)]+
    \text{tr}[\sigma_k \log_2(\rho_k)]|&\leq&
    |-\text{tr}[\sigma \log_2(\rho)] + \text{tr}[\sigma \log_2(\omega_k)]|\\
    &+&| - \text{tr}[\sigma \log_2(\omega_k)] + \text{tr}[\sigma_k \log_2(\omega_k)]|\nonumber\\
    &+&| - \text{tr}[\sigma_k \log_2(\omega_k)]+ \text{tr}[\sigma_k \log_2(\rho_k)]|,
    \nonumber
\end{eqnarray}
where $\omega_k:=
\|\sigma-\sigma_k\| \sigma_\beta
+ (1-\|\sigma-\sigma_k\|)
\rho$. Due to the operator monotonicity
of the logarithm
\begin{equation}
-\text{tr}[\sigma \log_2(\rho)] + \text{tr}[\sigma \log_2(\omega_k)]
\geq
\log(1- \|\sigma-\sigma_k\|)
\end{equation}
holds.
But $-\text{tr}[\sigma \log_2(\rho)]\leq - \text{tr}[\sigma \log_2(\omega_k)]$, and
therefore,
\begin{equation}
    \lim_{k\rightarrow\infty} |-\text{tr}[\sigma \log_2(\rho)]+
    \text{tr}[\sigma \log_2(\omega_k)]|=0.
\end{equation}
In the same way one finds that
$\lim_{k\rightarrow\infty} | - \text{tr}[\sigma_k \log_2(\omega_k)]+ \text{tr}[\sigma_k \log_2(\rho_k)]|=0$.
The third
term on the right hand side of Eq.\ (\ref{all})
can be dealt with just as in Ref.\ \cite{DonaldCont},
where the Gibbs state plays the
role of the maximally mixed state: Since
\begin{equation}
| - \text{tr}[\sigma \log_2(\omega_k)] + \text{tr}[\sigma_k \log_2(\omega_k)]|
\leq
\| \sigma - \sigma_k \|\,  |||  \log_2(\omega_k)     |||,
\end{equation}
one can again make use of the operator monotonicity of
the logarithm to find
\begin{equation}
|||  \log_2(\omega_k) |||\leq - \log_2(\|\sigma-\sigma_k\|)
+ ||| \log_2( \sigma_\beta)|||,
\end{equation}
and hence,
$\lim_{k\rightarrow\infty}| - \text{tr}[\sigma \log_2(\omega_k)] + \text{tr}[\sigma_k \log_2(\omega_k)]|=0$.
Collecting the partial results, one finds that
$\lim_{k\rightarrow\infty} |E_R(\sigma)-E_R(\sigma_k)|=0$.
\proofend

The corresponding asymptotic statement reads as follows:\\

\noindent
{\bf Proposition 9.} {\it
Let $\sigma\in {\cal S}_M({\cal H})$, $M>0$,
be a pure state that is supported
on a finite-dimensional
subspace of ${\cal S}({\cal H})$,
and let $\{\sigma_n\}_{n=1}^{\infty}$,
$\sigma_n\in {\cal S}_{n M}({\cal
H}^{\otimes n})$, be a sequence of states satisfying
$\lim_{n\rightarrow\infty}
    \|\sigma_n -\sigma^{\otimes n}\|=0$.
Then
\begin{equation}
    \lim_{n\rightarrow\infty}
    \frac{|E_R(\sigma^{\otimes n})
    - E_R(\sigma_n)|}{n}
    =0.\label{Asy}
\end{equation}
}

{\it Proof.} One may proceed as before. Again,
\begin{equation}
    |E_R(\sigma^{\otimes n})
    - E_R(\sigma_n)|
    \leq
    |S(\sigma^{\otimes n}) - S(\sigma_n)|
    +
    | - \text{tr}[\sigma^{\otimes n}\log_2(\eta_{n})+
    \text{tr}[\sigma_{n}\log_2(\rho_{n})
    ] |.
\end{equation}
By means of
Fannes'  inequality  one can infer that
$\lim_{n\rightarrow\infty} |S(\sigma^{\otimes n}) -
S(\sigma_n)| /n=0$. The second term on the right hand side
can be bounded from above as in Proposition 8, but now by
making use of
the $n$-fold product
$\sigma_{\beta}^{\otimes n}$ of the Gibbs state $\sigma_{\beta}$.
\proofend

This statement ends the considerations of entanglement measures in the
infinite-dimen\-sional setting. We have seen that if the mean energy
is bounded from above, and under an assumption concerning the spectrum
of the Hamiltonian $H$, then several entanglement measures retain their
trace-norm continuity.

\section{Non-existence of a separable ball}

Motivated by the findings of the two previous
sections one might be tempted to think that with
the help of a constraint on the mean energy, a
separable neighbourhood of some mixed state can
be recovered. In finite-dimensional bi-partite quantum systems,
the situation is as follows:
for any dimension of the Hilbert space
of a bi-partite system
there exists a separable neighbourhood
of the maximally mixed state  -- the
tracial-state
\cite{Clifton,Ball1}:
Whenever a state is closer to the maximally mixed state
with respect to the trace-norm (or any other norm),
then one
can be sure that the state is not entangled.
The size of this neighbourhood is however
not independent of the
dimension of the system: loosely speaking,
it decreases with
increasing dimension of the underlying Hilbert space.
In infinite-dimensional systems, the set of entangled
states is trace-norm dense in the state space of
the system, and there is no separable neighbourhood of
any mixed state \cite{Clifton}.
It is the purpose of the subsequent proposition to show
that also under the restriction that the mean energy is bounded from
above, no such neighbourhood can be reestablished.\\

\noindent
{\bf Proposition 10. } {\it
For any $\varepsilon>0$, $M>0$,
and  $\sigma\in {\cal S}_M({\cal H})$
there exists an entangled state
$\rho\in {\cal S}_M({\cal H})$
with the property
that $\| \sigma-\rho\|<\varepsilon$.}\\

{\it Proof. }\/
We will prove this statement by constructing a sequence
$\{\rho_k\}_{k=1}^{\infty}$ of entangled states
$\rho_k\in {\cal S}_M({\cal H})$ that satisfy
$\rho_k\longrightarrow \sigma$ in trace-norm.
Essentially, the idea is to construct a sequence
$\{\rho_{k}\}_{k=1}^{\infty}$
of states which converge to $\sigma$ in trace norm, but which are
entangled on a $2\times 2$-dimensional subspace.
Let $k\in{\mathbbm{N}}$,
\begin{equation}
 {\cal L}_{k}:=\text{span}  \bigl\{
    \phi_{A}^{(i)}\otimes \phi_{B}^{(j)} : i, j
    \in \{k,k+1\} \,
\bigr\},
\end{equation}
with $\phi_{A}^{(i)}$ and $\phi_{B}^{(j)} $
as in section 2,
and denote by $\pi_{k}$ the
projection on the Hilbert space
\begin{equation}
{\cal K}_{k}:=
\text{span}
\{\phi^{(l)}_A \otimes \phi^{(m)}_B:\,
l,m=0,\ldots, k-1\}.
\end{equation}
Let $\lambda_{k}:=\text{tr}[\pi_{k} \sigma\pi_{k}]$. If
$\lambda_{k}<1$,
then one can find a state vector
$\phi_{k}\in {\cal L}_{k}$
such that the partial transpose $(|\phi_{k}\rangle\langle\phi_{k}|)^{T_{A}}$
is not positive, and such that
\begin{equation}
    \text{tr}[H \sigma] - \text{tr}[H
\pi_{k}\sigma\pi_{k}]\geq (1-\lambda_{k})
\text{tr}[H
|\phi_{k}\rangle\langle\phi_{k}|].
\end{equation}
In this case set
\begin{equation}
    \rho_{k}:=\pi_{k}\sigma\pi_{k}+ (1-\lambda_{k})
|\phi_{k}\rangle\langle\phi_{k}|.
\end{equation}
If $\lambda_{k}=1$, then take $\rho_{k}:=
(1-1/k)\sigma + 1/k |\phi_{k}^{+}\rangle\langle\phi_{k}^{+}|$,
where $\phi_{k}^{+}:=(\phi_{A}^{(k)}\otimes \phi_{B}^{(k)}
+ \phi_{A}^{(k+1)}\otimes \phi_{B}^{(k+1)})/\sqrt{2}$.
The sequence $\{\rho_{k}\}_{k=1}^{\infty}$
has all the desired properties:
by definition $\rho_{k}\longrightarrow \sigma$ in
trace-norm, $\text{tr}[H \rho_{k}]\leq \text{tr}[H \sigma]<M$,
and $\rho_{k}$ is entangled for every $k\in{\mathbbm{N}}$:
the partial transpose of the projection of $\rho_{k}$ on
${\cal L}_{k}$ is not positive by construction, and hence,
this projection is entangled, according to the
Peres-Horodecki-criterion \cite{Peres}.
Therefore, $\rho_{k}$ is also an entangled
state.
\proofend


\section{Summary and Conclusion}

The content of this paper may be summarized in a nutshell as
follows: if one imposes a restriction on the mean energy and
requires that the Gibbs state exists, then several entanglement
measures are trace-norm continuous, just as in the
finite-dimensional setting. We investigated the entropy of
entanglement, the entanglement of formation, and the relative
entropy of entanglement, both for a single copy and in the
asymptotic limit. The findings are in a sense of less practical
than of reassuring nature: if the energy is bounded, then one may
truncate the state with respect to a finite dimensional subspace
and evaluate the degree of entanglement on this subspace. This
could be relevant in the situation where one considers a quantum
operation that acts on a finite-dimensional subspace of a quantum
system, which has nevertheless an infinite-dimensional Hilbert
space. An example of this type is the state manipulation of the
center-of-mass mode of ions in an ion trap, where one considers
only certain excitations. In an imperfect implementation of the
quantum operation the resulting state may not be strictly confined
to a finite-dimensional subset any more. But if the difference of
the actual state and the restriction on the finite-dimensional
subset is very small, one should not expect to have a completely
different situation as far as the entanglement of the state is
concerned, by taking the rest of the Hilbert space. In this sense
are the findings also relevant for investigations of the degree of
entanglement of Gaussian quantum states. Needless to say, there
are many issues that remain to be addressed. For example, 
the concepts of
entanglement of distillation, the `one-shot'-distillation
involving a single copy only, and the asymptotic entanglement cost
have so far only been formulated in the finite dimensional case in
a rigorous manner, and await a systematic investigation for
infinite-dimensional quantum systems.

\section{Acknowledgements}

We would like to thank K.\ ${\dot {\rm Z}}$yczkowski,
R.F.\ Werner and J.I.\ Cirac 
for valuable remarks.
This work has been supported by the European Union
(EQUIP -- IST-1999-11053),
the Alexander-von-Humboldt-Foundation,
the ESF, and the EPSRC.


\begin{thebibliography}{99}
\bibitem{asy}
In the finite dimensional setting the asymptotic continuity in this
sense -- together with convexity, normalization, and monotonicity
under local operations -- is required in order to  specify
the entropy of entanglement as the unique measure of entanglement
\cite{Unique}.
\bibitem{FannesIn}
    Let $\sigma$ and $\rho$ be states on a finite dimensional Hilbert
    space ${\cal H}$ with dimension $d$. Fannes' inequality \cite{Fannes,Ohya}
    states
    that if $\|\sigma-\rho\|<1/3$, then
    \begin{eqnarray}\nonumber
        |S(\sigma)-S(\rho)|\leq \log
        (d)
        \| \sigma-\rho\| -
        \| \sigma-\rho\| \log (\|
        \sigma-\rho\|).
    \end{eqnarray}
\end{thebibliography}
\end{document}